\documentclass[runningheads]{llncs}

\usepackage[T1]{fontenc}
\usepackage{amsmath}
\usepackage{hyperref}
  \usepackage{paralist}
  \usepackage{graphics} %% add this and next lines if pictures should be in esp format
  \usepackage{epsfig} %For pictures: screened artwork should be set up with an 85 or 100 line screen
  \usepackage[utf8]{inputenc}
\usepackage{graphicx}  \usepackage{epstopdf}%This is to transfer .eps figure to .pdf figure; please compile your paper using PDFLeTex or PDFTeXify.
 \usepackage{cases}
\usepackage{amssymb}
\usepackage{fancyhdr,graphicx,amsmath,amssymb}
\usepackage[ruled,vlined]{algorithm2e}
\include{pythonlisting}
 \usepackage{comment}
 \usepackage{tikz}
\usepackage{subcaption}
\usepackage{xcolor}
\usepackage{tikz} % To generate the plot from csv
\usepackage{subcaption}
\usepackage{subcaption}
\usepackage{tabularx,booktabs}
  \newcolumntype{Y}{>{\centering\arraybackslash}X}
\usepackage{array}
\usepackage{rotating}
\usepackage{graphicx}
\usepackage{tikz}
\usetikzlibrary{arrows.meta, 
                automata,
                intersections,
                positioning,
                quotes,
                matrix,
                bbox % <-- added
                }
\usetikzlibrary{shapes,arrows,fit}
\usetikzlibrary{graphs,quotes,arrows.meta}
\usepackage{forest}
\usetikzlibrary{arrows}
\usepackage{capt-of}
\forestset{
  direction switch/.style={
    for tree={
      if level=3{}{draw},
      thick,
      edge={thick},
      if level=1{
        child anchor=north,
        edge path={
          \noexpand\path [\forestoption{edge}] (!u.parent anchor) -- ++(0,-.5em) -| (.child anchor)\forestoption{edge label};
        },
        s sep+=.5em,
        for descendants={
          child anchor=west,
          align=left,
          edge path={
            \noexpand\path [\forestoption{edge}] (!u.parent anchor) ++(1em,0) |- (.child anchor)\forestoption{edge label};
          },
          fit=band,
        },
        for tree={
          parent anchor=south west,
          anchor=mid west,
          grow'=0,
          font=\sffamily,
          if n children=0{}{
            delay={
              prepend={[,phantom, calign with current]}
            }
          },
          before computing xy={
            l=2em
          },
        },
        before drawing tree={
          x/.wrap pgfmath arg={##1}{.6*x()},
          for children={
            x/.wrap pgfmath arg={##1+1em}{.6*x()},
            for children={
              x/.wrap pgfmath arg={##1+2em}{.6*x()},
            }
          }
        }
      }{
        if level=0{
          parent anchor=south,
          anchor=south,
        }{},
      },
    },
  }
}
\usetikzlibrary{matrix,chains,positioning,decorations.pathreplacing,arrows}
\newcommand{\arXiv}[1]{arXiv:\href{https://arxiv.org/abs/#1}{#1}}

\usepackage[T1]{fontenc}
\begin{document}
\title{Learning-Based Approaches for Job Shop Scheduling Problems: A Review }
\author{Karima Rihane \inst{1} \and
Adel Dabah \inst{2} \and
Abdelhakim AitZai\inst{1}}
\authorrunning{K. Rihane et al.}
\institute{University of Sciences and Technology Houari Boumediene
(USTHB), Algiers, Algeria \email{krihane@usthb.dz}; \email{h.aitzai@usthb.dz} \and
Forschungszentrum Jülich, Germany \\
\email{a.dabah@fz-juelich.de}\\
 }
\maketitle              % typeset the header of the contribution

\begin{abstract}
 Job Shop Scheduling (JSS) is one of the most studied combinatorial optimization problems. It involves scheduling a set of jobs with predefined processing constraints on a set of machines to achieve a desired objective, such as minimizing makespan, tardiness, or flowtime. Since it introduction, JSS has become an attractive research area. Many approaches have been successfully used to address this problem, including exact methods, heuristics, and meta-heuristics. Furthermore, various learning-based approaches have been proposed to solve the JSS problem. However, these approaches are still limited when compared to the more established methods. This paper summarizes and evaluates the most important works in the literature on machine learning approaches for the JSSP. We present models, analyze their benefits and limitations, and propose future research directions. \footnote{This paper has been accepted to appear at The International Conference on the Dynamics of Information Systems (DIS 2025)
June 1-5, 2025, London, UK.}
\keywords{Job Shop Scheduling Problems \and  Machine learning \and Artificial Neural
Network \and Reinforcement learning\and  Deep reinforcement learning}
\end{abstract}

\section{Introduction}
Scheduling problems are concerned with the allocation of a set of tasks (jobs) to a limited number of resources (machines) under certain constraints to optimize one or more objective functions. One of the most studied scheduling problems in the literature is the classical Job Shop Scheduling Problem (JSSP). This problem, which is among the NP-hard combinatorial optimization problems (COPs) \cite{nphard}, has an exponentially increasing search space. JSSP involves scheduling the processing of a finite set of $n$ jobs on a finite set of $m$ machines. Each job consists of $n_{i}$ operations, each representing the processing step of a job on a specific machine for an uninterrupted period. Each machine can handle one operation at a time. The classical objective of the JSSP is to find a schedule that minimizes the completion time of all jobs.

JSSP is a decision-making problem that appears in many real-world situations. Therefore, developing efficient algorithms to solve such problems has become an attractive area of research. Over the years, diverse approaches have been used to solve the JSSP. The first attempts to solve it to optimality were based on mixed-integer linear programming and Branch and Bound (B\&B) algorithms. However, due to the complex nature of the JSSP, these approaches can only solve small instances and remain inefficient for large problem instances. In this context, heuristics and meta-heuristics have been primarily used. These approaches make a trade-off between the quality of the solution and the time needed to obtain it. In other words, they return an acceptable solution quality in a reasonable time.

In the literature, we find the adaptation of many meta-heuristics to solve the JSSP. These include Genetic Algorithms (GA), Tabu Search (TS), Variable Neighborhood Search (VNS), Simulated Annealing (SA), Particle Swarm Optimization (PSO), and hybrid systems that combine heuristics and meta-heuristics. These methods have proven effective in solving the JSSP.

However, most real-world problems involve a large number of complex constraints, and adapting these classical approaches is still very challenging. In this context, there is a significant need for more robust and scalable approaches that can efficiently deal with the evolving complexity of real-world problems. One of the most active research areas today is the use of learning-based approaches for solving COPs via a data-driven approach. This helps in tackling the tremendous complexity of modern problems.

Machine Learning (ML) is a collection of Artificial Intelligence (AI) techniques that allow computers to learn and improve from data. Depending on how the learning is conducted, three sub-categories of ML can be distinguished: supervised learning, unsupervised learning, and reinforcement learning. In particular, Neural Networks (NN), Reinforcement Learning (RL), and the combination of these two approaches are the new trends in ML for addressing COPs.

This paper aims to summarize and analyze the most important contributions of neural networks and reinforcement learning approaches proposed for JSSPs. Furthermore, our study discusses the different techniques used and provides a comparison between them.

The remainder of the paper is structured as follows: Section \ref{sec:02} introduces the classical JSSP and the approaches used to solve it. Section \ref{sec03} describes learning-based approaches used to solve JSSPs, with a particular focus on NN and RL methods. Section \ref{sec:04} discusses and proposes insights on NN and RL models for JSSPs. Finally, conclusions are presented in Section \ref{sec:05}.

\label{sec:01} 

\section{Preliminaries }
\label{sec:02} 

\subsection{Formulation of the Problem}
The Job Shop Scheduling Problem is a classical combinatorial optimization problem in scheduling. It consists of determining the optimal time allocation of operations on a finite set of machines while satisfying all problem constraints. The primary objective is to minimize the makespan \(C_{max}\), which represents the maximum completion time of all jobs.

Formally, the JSSP has a finite set of  $n$  jobs $ J= \{    J_{1},J_{2},...,J_{n} \}$  that needs to be processed on a finite set of  $m$  machines $ M=  \{    M_{1},M_{2},...,M_{m}  \}$. 
Each job $J_{i}, i=1,...,n$ consists of $k$ operations  $  \left\{   O_{i1},O_{i2},...,O_{ik}  \right\}$, $k=1...,m$, each of them needs to be processed during an uninterrupted period $p$ on a given machine.
Each machine can handle at most one operation at a time. 
Let us consider $N$ the set of all operations $ N=  \{  O_{1},O_{2},...,O_{n.m}  \}$ . 
The objective is to find a schedule (feasible solution) that minimizes the makespan $C_{max}$, which represents the maximum completion time of all jobs.
The mathematical formulation is given as:
\begin{itemize}
\item \(t_{i}\): Start time of operation \(O_{i}\);
\item \(p_{i}\): Processing time of operation \(O_{i}\);
\item \(c_{i}\): Completion time of operation \(O_{i}\);
\item \(PC\): Set of precedence constraints between operations of the same job;
\item \(E_{k}\): Set of operations requiring processing on machine \(M_{k}\).
\end{itemize}

The constraints of the JSSP are as follows:
   \begin{numcases}
       \strut        $$ t_{i} + p_{i} \leq t_{j},  \forall ( O_{i}, O_{j}) \in PC $$ \label{e1} \\  
 $$(t_{j} - t_{i}) \geq p_{i}\  or\ (t_{i} - t_{j}) \geq p_{j} \forall(O_{i}, O_{j}) \in E_{k}, \forall k=  \left\{ 1...,m \right\}$$ \label{e2} \\
 $$ t_{i}\geq 0, \forall O_{i} \in N $$ \label{e3}
      \end{numcases}
      
Equation (\ref{e1}) represents the precedence constraints between the operations of the same job. Equation (\ref{e2}) enforces the sequencing constraints for operations that require processing on the same machine, ensuring no overlap in their execution times. Equation (\ref{e3}) indicates the earliest starting time of all operations.  

The classical objective is to minimize the makespan, \( C_{\text{max}} = \max\{c_i\}, \forall i \in \{1, \dots, n \cdot m\} \), which represents the maximum completion time of all operations. From a complexity perspective, the JSSP is known to be NP-hard \cite{nphard}, implying that the search space grows exponentially with the number of jobs and machines. This computational difficulty requires the use of advanced techniques to efficiently find feasible and near-optimal solutions.

The JSSP can also be represented using a disjunctive graph \cite{blazewicz2000}, a powerful graphical tool for modeling scheduling problems. A disjunctive graph, denoted as $G=(V,C,D)$, is a mixed graph where:
\begin{itemize}
   \item  $V$ represents the set of vertices, with each vertex corresponding to an operation. This includes two dummy vertices, start and end, both with zero processing time.
    \item $C$ is a set of directed arcs (conjunctions) that describe precedence constraints between operations within the same job.
    \item $D$ is a set of undirected edges (disjunctions) connecting pairs of operations that require the same machine, but whose execution order is not yet determined.       
\end{itemize}

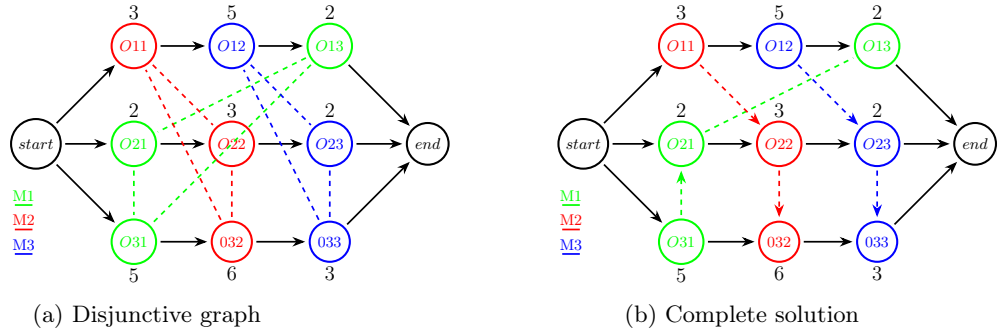
\begin{figure}[h]
%\centering

    \begin{subfigure}{0.3\textwidth}
       % \centering
       \resizebox{6cm}{!}{
        \begin{tikzpicture}[auto, node distance =20mm and 20mm,
        dot/.style = {circle, fill=black, inner sep=2pt, node contents={}},
        every state/.style = {circle, draw=black, very thick},
        every edge/.style = {draw=blue, line width=1pt, -Stealth,
                             shorten >=2pt, shorten <=2pt},
        every edge quotes/.style = {auto, sloped, inner sep=2pt},
        every label/.style = {rectangle, inner sep=2pt, font=\large},
        mincut/.style = {draw=green!40!black, line width=3pt, dashed}
                                              ]
   \node[state]  at (-1, 2)  (0) [black,label=above:{ $ $}] {$start$};
    \node[state]  at (1, 4)  (1) [ red,label=above:{ $3$}] {$O11$};
    \node[state]  at (3, 4)  (2) [blue,label=above:{ $5$}] {$O12$};
    \node[state]  at (5, 4)  (3) [green, label=above:{ $2$}] {$O13$};
    \node[state]  at (1, 2)  (4) [green,label=above:{ $2$}] {$O21$};
    \node[state]  at (3, 2)  (5) [red,label=above:{ $3$}] {$O22$};
    \node[state]  at (5, 2)  (6) [blue,label=above:{ $2$}] {$O23$};
    \node[state]  at (1, 0)  (7) [green,label=below:{ $5$}] {$O31$};
    \node[state]  at (3, 0)  (8) [red,label=below:{ $6$}] {$032$};
    \node[state]  at (5, 0)  (9) [blue, label=below:{ $3$}] {$033$};
    \node[state]  at (7, 2)  (T) [black,label=above:{ $ $}] {$end$}; 
    \path [bezier bounding box] % for improved bounding box calculation, requires bbox library
    (0)  edge["{}", black]                 (1)
    (0)  edge["{}", black]                 (4)
    (0)  edge["{}", black]                 (7)
    (1)  edge["{}", black]                 (2)
    (2)  edge["{}", black]                 (3)
    (3)  edge["{}", black]                 (T)
    (4)  edge["{}", black]                 (5)
    (5)  edge["{}", black]                 (6)
    (7)  edge["{}", black]                 (8)
    (8)  edge["{}", black]                 (9)
    (9)  edge["{}", black]                 (T)
    (6)  edge["{}", black]                 (T)
    % Arcs non dirigés (M1)
    (3)  edge[green, dashed, -]                (4) % Arc sans flèche
    (4)  edge[green, dashed, -]                (7) % Arc sans flèche
    (3)  edge[green, dashed, -]                (7) % Arc sans flèche

     % Arcs non dirigés (M2)
    (1)  edge[red, dashed, -]                (5) % Arc sans flèche
    (5)  edge[red, dashed, -]                (8) % Arc sans flèche
    (8)  edge[red, dashed, -]                (1) % Arc sans flèche
    
         % Arcs non dirigés (M2)
         
    (2)  edge[blue, dashed, -]                (6) % Arc sans flèche
    (6)  edge[blue, dashed, -]                (9) % Arc sans flèche
    (9)  edge[blue, dashed, -]                (2) % Arc sans flèche
%\node[state]  at (-1, 0.75)  (0) [rectangle, green,label=above:{ $$}] {$M1$};
(-1, 0.75) edge["{M1}",green, -] (-1.5, 0.75)
(-1, 0.25) edge["{M2}",red, -] (-1.5, 0.25)
(-1, -0.25) edge["{M3}",blue, -] (-1.5, -0.25);
%(-1, 0.75) edge["{M1}",blue, -] (-1.5, 0.75)
%\node[state]  at (-1, -0.25)  (0) [rectangle, red,label=above:{ $$}] {$M2$};
%\node[state]  at (-1, 0)  (0) [rectangle, blue,label=above:{ $$}] {$M3$};
\end{tikzpicture} 
}
        \caption{Disjunctive graph}
        \label{fig:sub1}
    \end{subfigure}
  \hspace{10em}
    \begin{subfigure}{0.4\textwidth}
     \resizebox{6cm}{!}{
        \begin{tikzpicture}[auto, node distance =20mm and 20mm,
        dot/.style = {circle, fill=black, inner sep=2pt, node contents={}},
        every state/.style = {circle, draw=black, very thick},
        every edge/.style = {draw=blue, line width=1pt, -Stealth,
                             shorten >=2pt, shorten <=2pt},
        every edge quotes/.style = {auto, sloped, inner sep=2pt},
        every label/.style = {rectangle, inner sep=2pt, font=\large},
        mincut/.style = {draw=green!40!black, line width=3pt, dashed}
                                              ]
    \node[state]  at (-1, 2)  (0) [black,label=above:{ $ $}] {$start$};
    \node[state]  at (1, 4)  (1) [ red,label=above:{ $3$}] {$O11$};
    \node[state]  at (3, 4)  (2) [blue,label=above:{ $5$}] {$O12$};
    \node[state]  at (5, 4)  (3) [green, label=above:{ $2$}] {$O13$};
    \node[state]  at (1, 2)  (4) [green,label=above:{ $2$}] {$O21$};
    \node[state]  at (3, 2)  (5) [red,label=above:{ $3$}] {$O22$};
    \node[state]  at (5, 2)  (6) [blue,label=above:{ $2$}] {$O23$};
    \node[state]  at (1, 0)  (7) [green,label=below:{ $5$}] {$O31$};
    \node[state]  at (3, 0)  (8) [red,label=below:{ $6$}] {$032$};
    \node[state]  at (5, 0)  (9) [blue, label=below:{ $3$}] {$033$};
    \node[state]  at (7, 2)  (T) [black,label=above:{ $ $}] {$end$}; 
    \path [bezier bounding box] % for improved bounding box calculation, requires bbox library
    (0)  edge["{}", black]                 (1)
    (0)  edge["{}", black]                 (4)
    (0)  edge["{}", black]                 (7)
    (1)  edge["{}", black]                 (2)
    (2)  edge["{}", black]                 (3)
    (3)  edge["{}", black]                 (T)
    (4)  edge["{}", black]                 (5)
    (5)  edge["{}", black]                 (6)
    (7)  edge["{}", black]                 (8)
    (8)  edge["{}", black]                 (9)
    (9)  edge["{}", black]                 (T)
    (6)  edge["{}", black]                 (T)
    % Arcs non dirigés (M1)
    %(3)  edge[green, dashed, -]                (4) % Arc sans flèche
    (4)  edge["{}",green, dashed, -]                (3) % Arc sans flèche
    (7)  edge["{}",green, dashed,]                (4) % Arc sans flèche

     % Arcs non dirigés (M2)
    (1)  edge["{}",red, dashed,]                (5) % Arc sans flèche
    (5)  edge["{}",red, dashed,]                (8) % Arc sans flèche
   % (8)  edge[red, dashed, -]                (1) % Arc sans flèche
    
         % Arcs non dirigés (M2)
         
    (2)  edge["{}",blue, dashed,]                (6) % Arc sans flèche
    (6)  edge["{}",blue, dashed,]                (9) % Arc sans flèche
  %  (9)  edge[blue, dashed, -]                (2) % Arc sans flèche
(-1, 0.75) edge["{M1}",green, -] (-1.5, 0.75)
(-1, 0.25) edge["{M2}",red, -] (-1.5, 0.25)
(-1, -0.25) edge["{M3}",blue, -] (-1.5, -0.25);

\end{tikzpicture} 
}
        \caption{Complete solution}
        \label{fig:sub2}
       
    \end{subfigure}

    \caption{Disjunctive graph representation for JSSP. (a) Visualizes the JSSP instance described in Table \ref{tab:01}, where black arrows represent conjunctive arcs (precedence constraints), and dotted lines denote disjunctive arcs grouped into machine-specific cliques, each distinguished by a unique color. (b) Depicts a complete solution where all disjunctive arcs have been assigned directions, resulting in a directed acyclic graph}
    \label{fig:main}

\end{figure}

Consequently, finding a solution to a JSSP instance is equivalent to fixing the
direction of each disjunction, such that the resulting graph is a Directed Acyclic Graph. An example of disjunctive graph for a JSSP instance in Table \ref{tab:01} and its solution are depicted in Figure \ref{fig:main} (a) and (b), respectively.

%Table \ref{tab:01} represents BJSS instance with two jobs and three machines. The first job ($J1$) has 5 min processing time on machine $M1$, 3 min on $M2$, and 8 min on machine $M3$. The second job ($J2$) has 8 min processing time on machine $M2$, 2 min on $M1$, and 7 min on machine $M3$.
\begin{table}[h]
\renewcommand{\thetable}{\arabic{table}}
 \centering
\caption{JSSP instance with three jobs and three machines.}
The (3*9) matrix contains two (3*3) matrices. The first matrix represents the processing times of the two jobs on the 3 machines.\\
\begin{tabular}{ccc}
\hline\noalign{\smallskip}
job & sequence & processing times \\
\noalign{\smallskip}\hline\noalign{\smallskip}
$J1$ & $M2, M3, M1 $& 3, 5, 2 \\
$J2$ & $M1, M2, M3 $& 2, 3, 2 \\
$J3$ & $M1, M2, M3 $& 5, 6, 3 \\
\noalign{\smallskip}\hline
%%\label{tab:01}
\end{tabular}
\label{tab:01}
\end{table}
%%%%%%%%%%%%%%%%%%%%%%%%%%%%%%%%%

\subsection{Job-Shop Scheduling Problem and Resolution Techniques}

Since the early efforts to solve the JSSP, researchers have employed a variety of techniques, which can be classified into two main categories: exact and approximate methods.

Exact methods guarantee an optimal solution for the JSSP. These include algorithms such as A-star, Linear Programming, Dynamic Programming, and Branch-and-Bound methods. These approaches have been extensively utilized for optimal solutions of the JSSP \cite{23,39,53}. However, due to the NP-hard nature of the JSSP, these methods are only effective for small-scale instances. Consequently, approximate methods are employed to handle larger-scale benchmarks.

Approximate methods strike a balance between time complexity and solution quality. They can be further divided into heuristics and metaheuristics. Various heuristics have been proposed to solve the JSSP, often based on dispatching rules \cite{16,40} and the Shifting Bottleneck procedure \cite{17,38,74}. While these heuristics offer favorable time complexity, they tend to produce lower-quality solutions and are often tailored to specific instances, making them less adaptable to other variants.

In contrast, metaheuristics typically yield high-quality solutions in a relatively short runtime and are versatile enough to be applied to a broad range of optimization problems. Population-based metaheuristics such as Particle Swarm Optimization \cite{64,75}, Genetic Algorithms \cite{1,13,15,genticN}, Artificial Immune Systems \cite{2}, and Hybrid Systems \cite{56,96} have been successfully used to solve the JSSP. Additionally, local search metaheuristics, including Tabu Search \cite{19,55}, Variable Neighborhood Search, and Simulated Annealing \cite{29,76}, have shown to require less computational effort and provide robust solutions under varying constraints.

The evolving nature of real-world scheduling problems calls for the development of more powerful and adaptive techniques. This has led researchers to explore AI-driven approaches, particularly ML-based approaches, to address the JSSP through data-driven methods. The following section will examine and analyze the application of ML techniques to solve the JSSP.

\section{Machine Learning Approaches for Job Shop Scheduling Problem }
\label{sec03}
Machine learning is a subfield of artificial intelligence that enables computers to learn and improve their performance without being explicitly programmed. Traditionally, machines execute tasks based on instructions provided by programmers. However, in the ML paradigm, machines learn from past experiences and data by analyzing patterns, identifying changes, and adapting themselves to achieve specific tasks.

Based on the learning approach, ML can be categorized into three main subfields: supervised learning, unsupervised learning, and reinforcement learning. %Figure \ref{fig:01} provides an overview of these ML subfields and their associated algorithms.

\textbf{Supervised Learning}: This type of ML involves using labeled datasets containing both input features and their corresponding response values to train a model. The model learns to predict response values for new, unseen data based on the training examples.

\textbf{Unsupervised Learning}: In contrast to supervised learning, this approach does not rely on labeled data. Instead, the model analyzes the data to uncover hidden patterns, structures, and features autonomously.

\textbf{Reinforcement Learning}: This is a dynamic programming-based approach where an agent interacts with its environment to learn through trial and error. By observing the outcomes of its actions, the agent adjusts its strategy to maximize a predefined reward function.

Over the years, ML approaches have achieved significant success in various fields, including speech and image recognition, gaming, object detection, and natural language processing. However, ML methods still face challenges when applied to COPs. Effectively leveraging ML to solve COPs remains an active area of research. The primary challenge lies in adapting ML techniques to learn from COP-generated data and generalizing this learning to address similar problems.

In the literature, numerous ML-based methods have been proposed to tackle various categories of COPs, as reported in \cite{Vesselinova2020,Mazyavkina2021,Cappart2021_arxiv}, \cite{Peng2021,Cappart2021_IJCAI,Zhang2023,Souza2024}. Among these learning-based approaches, neural networks and reinforcement learning have emerged as the most popular techniques. NNs offer significant strengths for COPs, they excel at capturing complex patterns, scaling to high-dimensional data, and generalizing to unseen instances. 
NNs are flexible, with architectures adaptable to various COPs, and provide rapid solution generation once trained, making them ideal for real-time applications. 
They integrate well with other methods, enhancing heuristics and guiding search spaces, and are robust to uncertainty and noise. 
Additionally, NNs enable end-to-end learning frameworks, incorporating problem-specific constraints directly into their design, which is particularly advantageous for complex problems combinatorial optimization problem.

RL offers several advantages when applied to COPs, it is particularly well-suited for solving problems that involve sequential decision-making, as it learns optimal policies through interactions with the environment. RL can handle dynamic and uncertain scenarios, making it effective for real-world COPs where problem parameters may change. Its ability to explore large and complex solution spaces allows it to discover high-quality solutions that might be difficult to find using traditional methods. Moreover, RL can adapt to different COPs by leveraging reward signals to guide the optimization process, offering a flexible and scalable approach to tackling diverse challenges.

The JSSP is a prominent example of a COP with significant real-world applications. Among the various approaches, neural networks and reinforcement learning have emerged as promising techniques for tackling its complexities. In the following sections, we will explore the application of these methods to JSSPs and analyze their potential impact on solving such problems.

\subsection{Neural Networks}

The functioning of biological neural networks such as parallelism, learning, adaptation, and generalization has inspired researchers to develop artificial neural networks that mimic the workings of the human brain. In 1943, McCulloch and Pitts \cite{87} introduced the first mathematical formulation of an artificial neuron. Later, in 1960, Rosenblatt \cite{60} developed the basic neural network model, introducing the three-layer perceptron. Since then, artificial NNs have become a highly active area of research.
\begin{figure}[h!]
\begin{tikzpicture}[
init/.style={
  draw,
  circle,
  inner sep=2pt,
  font=\Huge,
  join = by -latex
},
squa/.style={
  draw,
  inner sep=2pt,
  font=\Large,
  join = by -latex
},
start chain=2,node distance=13mm
]
\node[on chain=2] 
  (x2) {$x_2$};
\node[on chain=2,join=by o-latex] 
  {$w_2$};
\node[on chain=2,init] (sigma) 
  {$\displaystyle\Sigma$};
\node[on chain=2,squa,label=above:{\parbox{2cm}{\centering Activate \\ function}}]   
  {$f$};
\node[on chain=2,label=above:Output,join=by -latex] 
  {$y$};
\begin{scope}[start chain=1]
\node[on chain=1] at (0,1.5cm) 
  (x1) {$x_1$};
\node[on chain=1,join=by o-latex] 
  (w1) {$w_1$};
\end{scope}
%%%%%1
\begin{scope}[start chain=3]
\node[on chain=3] at (0,-1.25cm) 
  (.) {.};
\node[on chain=3,label=below:%Weights
%join=by o-latex
] 
  (.) {.};
\end{scope}
%%%%%
\begin{scope}[start chain=3]
\node[on chain=3] at (0,-1cm) 
  (.) {.};
\node[on chain=3,label=below:%Weights
%join=by o-latex
] 
  (.) {.};
\end{scope}
%%%%
\begin{scope}[start chain=3]
\node[on chain=3] at (0,-0.75cm) 
  (.) {.};
\node[on chain=3,label=below:%Weights
%join=by o-latex
] 
  (.) {.};
\end{scope}
%%%%%%
\begin{scope}[start chain=3]
\node[on chain=3] at (0,-0.5cm) 
  (.) {.};
\node[on chain=3,label=below:%Weights
%join=by o-latex
] 
  (.) {.};
\end{scope}
\begin{scope}[start chain=3]
\node[on chain=3] at (0,-1.5cm) 
  (xn) {$x_N$};
\node[on chain=3,label=below:Weights,join=by o-latex] 
  (wn) {$w_N$};
\end{scope}

\node[label=above:\parbox{2cm}{\centering Bias \\ $b$}] at (sigma|-w1) (b) {};

\draw[-latex] (w1) -- (sigma);
\draw[-latex] (wn) -- (sigma);
\draw[o-latex] (b) -- (sigma);

\draw[decorate,decoration={brace,mirror}] (x1.north west) -- node[left=10pt] {Inputs} (xn.south west);
\end{tikzpicture}
\caption{Model of neuron in network.} \label{fig:02}
\end{figure}
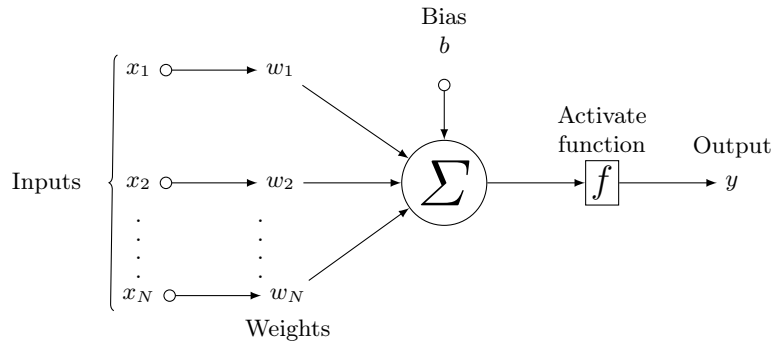

A neural network consists of small units called "neurons," which are organized into multiple layers. Neurons in one layer interact with neurons in the next layer through "weighted connections," which are real-valued. A neuron receives the values from connected neurons $(x_1,x_2,...,x_N)$ and multiplies them by their respective connection weights $(w_1,w_2,...,w_N)$. The sum of all connected inputs, along with the neuron’s bias $b$, is then passed through an "activation function" $f$, which mathematically transforms the value before it is transmitted to the next layer of neurons. In this way, the inputs are propagated throughout the entire network. The goal is to learn how to adjust the weights to achieve the desired output $y$.
Additionally, Figure \ref{fig:02} illustrates the general model of a neuron.
Since the introduction of the first neural network model, many variations have been developed. These models differ in their activation functions, learning algorithms, and architectures.  Well-known activation functions, learning algorithms, and neural network architectures are reported in Table \ref{table:2}.

\begin{table}[ht]
\renewcommand{\thetable}{\arabic{table}}
\centering
\caption{Well-known activation functions, learning algorithms, and neural network architectures.}
\begin{tabular}{c >{\raggedright\arraybackslash}p{8cm}}  % La deuxième colonne a une largeur définie
\hline\noalign{\smallskip}
Characterization & Examples\\
\noalign{\smallskip}\hline\noalign{\smallskip}
Activation Functions & Sigmoid, Tanh, ReLU (Rectified Linear Unit), Leaky ReLU, Softmax, Swish, ELU (Exponential Linear Unit), GELU (Gaussian Error Linear Unit), Softplus, Hard Sigmoid. \\ \hline
Learning Algorithms & Gradient Descent, Stochastic Gradient Descent (SGD), Adam, RMSprop, Adagrad, AdaDelta, Momentum, Nesterov Accelerated Gradient (NAG), L-BFGS. \\ \hline
Neural Network Architectures & Feedforward Neural Networks (FNN), Convolutional Neural Networks (CNN), Recurrent Neural Networks (RNN), Long Short-Term Memory (LSTM), Gated Recurrent Unit (GRU), Autoencoders, Variational Autoencoders (VAE), Generative Adversarial Networks (GANs), Transformer Networks, Capsule Networks, Deep Belief Networks (DBN), Radial Basis Function Networks (RBFN). \\
\noalign{\smallskip}\hline
\end{tabular}

\label{table:2}
\end{table}

Numerous artificial neural network models have been proposed to tackle job shop scheduling problems. The primary distinctions between these models arise from the choice of activation functions, network architectures, and learning algorithms used. In the following, we will review the literature on job shop scheduling problems addressed using NNs, highlighting the various approaches and their effectiveness.

\subsubsection{Shallow Neural Networks}

Hopfield and Tank \cite{34} introduced a deterministic neural network model capable of solving constraint satisfaction and optimization problems by translating the problem into units with predefined fixed-weighted connections. The Hopfield Neural Network (HNN) is a Recurrent Neural Network (RNN) organized in a single layer of neurons, with each neuron representing an element of the constraint optimization problem matrix. The network uses a sigmoid activation function to calculate the neuron outputs and adjusts the weights of its connections using a target function equivalent to the problem’s objective function to reach a stable state (solution). The HNN was initially used to solve the Traveling Salesman Problem (TSP) and Linear Programming (LP), establishing the HNN as an optimization technique.

Foo and Takefuji \cite{72,73} formulated the JSSP as a mixed integer linear program and adapted the HNN model with integer adjustments. However, this model led to constraints violations, resulting in infeasible solutions. To address these issues, Van Hulle \cite{47} modeled JSSP as a goal programming problem, mapping it to a goal programming network that guarantees feasible solutions through a relaxation strategy and binary adjustments until convergence. Despite this improvement, the search space remained too large and convergence was slow. To improve convergence and reduce the search space, Willems and Rooda \cite{77} proposed a new neural network structure with feedback connections to map the integer linear representation of the JSSP. Their approach aimed to: (i) reduce the search space by setting a minimum starting time (thresholds) for each operation and (ii) eliminate the need for integer adjustments during the calculation. This NN structure ensures rapid convergence to a feasible and optimal solution, although the thresholds used may sometimes lead to convergence toward a sub-optimal solution. To control the evolution of the model, force convergence to a final solution, and ensure the feasibility of solutions, Sabuncuoglu and Gurgn \cite{30} added an external processor to the HNN. This model was applied to solve the JSSP. 

In Hopfield Neural Networks, the choice of initial starting times significantly influences convergence and solution quality. Rukhsana and Sache \cite{46} addressed this issue by proposing a modified HNN that uses a tailored heuristic for initializing starting times. They further expanded the search space to enable the network to explore both optimal and near-optimal solutions.

Yang and Wang \cite{67} mapped the job shop scheduling problem to a constraint satisfaction adaptive neural network model (CSANN). In this model, the weights and biases are adjusted dynamically to produce feasible solutions. They also proposed three heuristics integrated with CSANN: the first heuristic aims to accelerate the process and ensure feasible solutions, the second targets achieving the expected makespan, and the third focuses on improving the quality of the obtained solutions. The results demonstrated that these three hybrid approaches significantly enhanced performance and solution quality compared to standalone CSANN solutions. In their subsequent work \cite{68}, Yang and Wang introduced a new hybrid heuristic to their earlier model to derive a no-delay schedule from the feasible solutions generated by CSANN. Later, Yang \cite{65} enhanced the adaptive neural network (CSANN) from \cite{67}, resulting in a new version called CSANN-II. The key improvement in CSANN-II was the integration of an adaptive RC-block (Resource Constraints units) construction. This mechanism dynamically adjusts to reflect the actual resource constraint satisfaction conditions during the neural network's execution. Additionally, CSANN-II was combined with the heuristics proposed in \cite{67}. The performance of CSANN-II and its hybrid approaches was benchmarked against the results presented by Giffler and Thomson \cite{23}, demonstrating superior performance and solution quality. Further, Yang \cite{66} combined CSANN-II with a local search approach to solve the JSSP. In this hybrid approach, the feasible solution obtained by CSANN-II is relaxed using a schedule relaxation technique to create a partially relaxed schedule. Starting from the relaxed solution, the local search iteratively swaps two concurrent jobs on each machine to generate a set of neighboring solutions. The algorithm explores these neighboring solutions, moving from one to another, until it converges to an optimal or near-optimal solution. This hybrid approach improved the performance of CSANN-II, particularly in terms of solution quality.

Lagrangian relaxation, a robust technique for separable integer programming problems, was employed to relax the JSSP formulation using Lagrange multipliers. This relaxation transformed the JSSP into separable sub-problems, Luh et al.\cite{54} proposed a RNN model to solve each sub-problem. The results of this hybrid method outperformed existing neural network-based approaches in the literature. 

In contrast to mathematical programming-based formulations, El-Bouri and Shah \cite{3} proposed a classifier neural network to enhance priority dispatching rules for solving the JSSP. Their approach aimed to improve the performance of traditional dispatching rules by using the neural network to predict the most suitable rule for each machine in the scheduling process. While this method demonstrated superior performance compared to conventional dispatching rules, it incurred a high computational cost due to the backpropagation-based training process.

Weckman et al. \cite{22} introduced another job shop scheduler based on a supervised multi-layer perceptron neural network. The network was trained to extract knowledge from optimal solutions generated by a genetic algorithm, capturing common features across these solutions. Once trained, the model was capable of generating schedules for new problem instances. The approach, tested against GA and traditional dispatching rules, demonstrated its ability to produce high-quality solutions, including optimal and near-optimal schedules. Similarly, Chaudhuri and De \cite{5} proposed a scheduler based on a rough fuzzy multi-layer perceptron neural network. This model aimed to learn common patterns in optimal solutions and generate high-quality schedules for new problems. By integrating rough and fuzzy sets, the model addressed the uncertainty and imprecision inherent in scheduling problems. The rough fuzzy classifier effectively produced strong solutions by leveraging the learned characteristics of GA-derived solutions. Telchy and Raafat \cite{20} took a different approach by using a three-layer feedforward backpropagation neural network (FFBNN) to prioritize and determine the starting order of each operation within the JSSP. Their model focused on establishing priorities and sequencing operations to improve the overall scheduling process.

Sim et al. \cite{50} proposed a multi-layer perceptron scheduler to minimize the makespan of our classical JSSP. The model follows the logic of the NN implemented in \cite{22}. It extracts features from GA-generated schedules and learns a priority class for each job operation. The NN breaks the trained data into twelve features representing operations position, processing time, and machine load. The network is composed of an input layer, two hidden layers, and an output layer. The input layer consists of twelve neurons, one neuron per feature. While the output layer is made up of six neurons, each neuron represents the operation priority class. During the process, the model may predict the same priority to two operations in the same machine; in this case, \cite{22} chooses arbitrarily. Conversely,  the model in \cite{50} uses dispatching rules to break these ties. The NN was trained using $6\times 6$ job shop schedules and tested for several benchmark instances. 

Combined with the PSO method, Zhang et al. \cite{97} proposed an error feedback neural network (NN). This network consists of four layers, including two hidden layers. The input and output layers are similar to those used in \cite{50} and \cite{22}, while the other two layers are the hidden layers. The model introduces a formula to determine the number of neurons in each layer. The PSO method is utilized to optimize the NN weights instead of employing a genetic algorithm (GA) for training the NN. In this approach, the position of each particle in the swarm represents the connection weights of the NN. At the start of the process, the particles are initialized randomly. During each iteration, the PSO algorithm searches for the optimal position, after which the NN updates its connection weights. By applying the optimization rules of PSO, the NN minimizes the learning error, calculated as the difference between the predicted and expected outputs.

So far, we have explored various types of NN models designed for JSSPs, summarized in Table \ref{tab nn}. These models have demonstrated their efficiency in solving constraints satisfaction problems and identifying appropriate dispatching rules. Additionally, they have proven effective in both supervised and unsupervised learning scenarios, with particularly strong performance when hybridized with other approaches. However, most NNs used for JSSPs rely on shallow architectures, typically consisting of only one or two hidden layers. This limitation can reduce the network's accuracy and hinder its ability to generalize. Furthermore, these shallow models often struggle to extract sufficient features from the data, leading to accuracy issues. To address these challenges, it would be more suitable to adapt Deep Learning (DL) techniques to JSSPs. 

\begin{table}[h!]
\renewcommand{\thetable}{\arabic{table}}
 \centering

\caption{Shallow Neural Networks models for JSSPs.}
 \resizebox{13.5cm}{!}{
\begin{tabular}{c|c|c}

\hline\noalign{\smallskip}
Study & NN Model & JSSP representation  \\
\noalign{\smallskip}\hline\noalign{\smallskip}
Hopfield and Tank \cite{34} &Hopfield NN & Constraint satisfaction problem \\
Foo and Takefuji,  \cite{72,73} &Adapted Hopfield NN & Mixed integer linear programming \\
Van Hulle \cite{47} &Goal programming network& Mixed integer linear programming \\
Willems and Rooda \cite{77} & Feedback network structure& Integer linear programming formulation \\
Sabuncuoglu and Gurgn \cite{30} &Adapted Hopfield NN& Constraint satisfaction problem\\
Rukhsana and Sache \cite{46} &HNN combined with a constructive heuristic & Constraint satisfaction problem\\
Yang and Wang \cite{67,68,66} &Adaptive networks guided by constructive heuristics & Constraint satisfaction problem\\
Luh et al. \cite{54} &RNN combined with Lagrangian relaxation method &Separable integer programming problem\\ 
El-Bouri and Shah \cite{3} & Classifier Neural Network & Dispatching rule-based model\\
Weckman et al. \cite{22}  &Supervised Multi-Layer Perceptron NN& GA-derived schedules\\
Chaudhuri and De \cite{5} &Rough Fuzzy Multilayer Perceptron NN& GA-derived schedules\\
Telchy and Raafat \cite{20} & Feedforward backpropagation NN& Dispatching rule-based model \\
Sim et al. \cite{50} & Multi-layer perceptron NN & GA-derived schedules\\
Zhang et al. \cite{97} & Error feedback neural network& PSO-derived schedules\\
\noalign{\smallskip}\hline
%%\label{tab:01}

\end{tabular}
}
\label{tab nn}

\end{table}

\subsubsection{Deep Neural Networks}

Deep learning, as introduced by Hinton et al. in \cite{hinton2006fast}, refers to the use of Deep Neural Networks (DNNs), which are designed to automatically learn hierarchical patterns and representations from large volumes of data. The 'depth' refers to the number of layers in the network, enabling the extraction of increasingly abstract features at each layer. This makes DNNs particularly powerful for applications such as image recognition, natural language processing, and speech recognition, where traditional neural networks might struggle to achieve high performance. Thus, deep learning marks a significant evolution in neural network models, pushing the boundaries of what is possible in machine learning and artificial intelligence. The ability of DNNs to learn complex, hierarchical representations of scheduling data opens up new possibilities for tackling the combinatorial nature of JSSPs with higher efficiency and accuracy. 

Zang et al. \cite{95} proposed a hybrid deep NN scheduler, it combines a deep convolution network with a back propagation network. The model, like \cite{22}, attempts to extract characteristics from GA-generated training samples. The model separates the JSSP instances into sub-problems based on job processing and machine priority at the training step. Then it applies a convolution two-dimensional transformation. It is a Cartesian product that converts each operation's one-dimensional features into a two-dimensional correlation between features. The 1-D and 2-D features form the inputs of the NN, named input1 and input2. They use the L1 full connected layer and L1 convolutional layer, respectively. L1 and L2 are two fixed numbers. The model includes a Flatten layer that combines all the inputs. The model uses the L2 full connected layer and passes it into the output layer. The output layer consists of the priority class of each job. The authors employ the error backpropagation approach to train the network parameter. The approach shows a good classification accuracy for various JSSP instances in minimizing the makespan. However, the architecture of this deep network is complicated and requires a long training time. In addition, the proposed convolution two-dimensional transformation approach has no well-defined rules for choosing parameters L1 and L2. 

Selection hyper-heuristics determine the best heuristic to apply during the solving process of JSSPs. Lara-Cárdenas et al. \cite{18} integrated this approach with a multilayer perceptron (MLP) neural network to minimize the makespan for of sixty $15 \times 15$ JSSP. The MLP processes six input features—three characterizing the schedule and three describing the problem state—to optimize heuristic selection. The network's architecture includes three or four hidden layers (tested in three typologies), with six output neurons corresponding to six well-known dispatching rules. While the approach yielded promising results, the authors did not establish a concrete method for determining the optimal number of hidden layers and neurons. Training relies on simulated annealing to generate hyper-heuristics, but this process is computationally intensive and sensitive to parameter tuning. The NN effectively adjusts weights to enhance the selection process but highlights challenges in balancing efficiency and learning complexity.

For large-scale and complex JSSPs, Shao and Kim \cite{88} adapted a deep learning-based method. The approach uses GA-generated solutions of $10 \times 10$ JSSP instances to learn scheduling decisions for minimizing the makespan of new instances. The model divides the training data into two categories of features: detailed-level and system-level. The authors define eighteen variables representing detailed-level information based on the processing times of jobs on machines. Meanwhile, the K-means method is employed to extract system-level features. To improve model precision, feature extraction at the detailed level must consider time steps, which requires significant storage space. To address this, the model employs two-layered Long Short-Term Memory (LSTM) networks.

LSTM, a type of recurrent neural network, is specifically designed to alleviate the long-term dependency issues of traditional RNNs, enabling long-term data storage \cite{63}. The model adds a concatenate layer to fuse the detailed-level and system-level features. The \cite{88} approach produces two types of output. The first is the reconstructed input state, and the second is an $m$-target node output, where each node corresponds to a machine's priority. For reconstructing input states, the model includes two symmetrical LSTM layers with self-supervised learning and one dense layer with $m$ nodes to predict machine priorities as the target output. To adjust its weights, the neural network calculates a loss function by combining two components using predefined parameters $\alpha$ and $\beta$. The first component, the mean squared error (MSE), is responsible for reconstructing input states, while the second, the categorical cross-entropy function, predicts the target output. The model is tested by calculating the makespan of unseen JSSP instances. Compared with \cite{95}, the results demonstrate the model's robustness in solving complex instances. The tests reveal that the makespan is heavily influenced by the loss function parameters. Additionally, the dual-channel architecture (detailed-level and system-level) complicates the training and learning phases.

Optimization methods typically require deep knowledge of the problem to achieve fast convergence and low execution times. However, such prior knowledge is often unavailable for real-world problems. deep neural netwoks approaches offer automatic feature discovery, reducing the reliance on domain expertise in solving complex problems. For instance, in a constrained optimization problem where the objective function value is already known, the model can incorporate the objective function as an equality constraint. This reduces the size of the search space and accelerates the solving process.

Wang et al. \cite{81} aimed to accelerate the resolution process of the JSSP by using a DNN to predict the makespan and then solving it based on the predicted value. The JSSP is represented as a 2D matrix containing the required machine for each job, the precedence dependencies between a job's operations, and the processing time of each operation. The proposed DNN is a CNN composed of two parts. The first part consists of a stack of convolutional layers that extract high-level abstractions from the input 2D matrix, capturing its essential features. The second part comprises fully connected layers that predict the makespan value. To mitigate overfitting during training, the model incorporates a dropout layer. This dropout mechanism leads to the prediction of multiple makespan values for each JSSP instance, and the final output is obtained by averaging these predictions. Then a designed strategy search takes the predicted makespan and solves the JSSP instance. 
The DNN is trained using $9 \times 9$ JSSP instances randomly generated and solved optimally by a constraints satisfaction problem. Once the model is trained, it is tested on benchmark instances. The results show that the proposed deep network accelerated the solving process.

The JSSP, formulated as a mixed-integer linear programming problem, can be solved optimally using a B\&B algorithm, which involves two sequential decision-making steps: node selection and variable selection. The strong branching rule is the most effective branching technique for constraint satisfaction problems, focusing on variable selection. However, it is computationally expensive. To address this, Juros et al. \cite{33} aim to accelerate the execution time of B\&B by employing a deep network model that mimics the strong branching rule. The proposed model takes the states of the B\&B algorithm as inputs and outputs an approximation of the variable selected by the strong branching rule. These states are represented as a bipartite graph, which justifies the use of a Graph Neural Network (GNN), a type of neural network specifically designed to process graph-structured data \cite{graphnn}. The bipartite graph consists of constraint nodes, variable nodes, and edges that connect the nodes whenever a constraint refers to a variable. At each step of the B\&B process, the states are represented by the bipartite graph, parameterized by node and edge feature matrices. The feature topology highlights the need for a convolutional layer in the proposed deep neural network, which is well-suited to processing data with a grid-like topology. The network is trained using $10 \times 10$ JSSP instances that are randomly generated and solved using the B\&B algorithm with the strong branching policy. During training, the input data passes through the graph convolution layers, which produce a learned probability distribution for each candidate branching variable. The test results show that the network successfully learns the strong branching policy, particularly for small instances, where the model outperforms the baseline solver. However, for larger instances, the model demonstrates some shortcomings, occasionally failing to select the correct branching variable and steering the search toward suboptimal regions of the search space. This behavior suggests that the model requires additional training with more data samples, which is both time- and resource-intensive. Alternatively, it is possible that the parametrization of the states failed to adequately capture the desired knowledge. %The 
 
Wang et al. \cite{GNN1} investigate the application of GNNs and graph transformer models to tackle COPs, with a particular focus on the JSSP. The model functions in two main steps; First, Representation Learning: The model first processes the JSSP instance using a GNN, which takes the graph representation of the problem as input. The GNN layers aggregate information from neighboring nodes (jobs, operations, machines) to generate node embeddings. These embeddings capture the relationships and dependencies within the scheduling problem. A pooling operation then combines these node embeddings into a single graph-level embedding. Second, Prediction: The graph-level embedding is passed to a functional component, like a Multilayer Perceptron, which predicts the optimal makespan of the schedule based on the learned graph representation. 

Corsini et al. \cite{GNN2} proposed a model that tackles the JSSP by framing it as a sequence of decisions, represented as a disjunctive graph, leveraging a generative encoder-decoder architecture inspired by the Pointer Network \cite{PointerNet} (a well-known encoder-decoder architecture for generating sequences of decisions). The encoder captures instance-wide relationships by transforming raw operation features into embeddings that incorporate the graph's topological information. This is achieved using two stacked Graph Attention Network (GAT) layers (a type of GNN), which leverage the disjunctive graph structure to enrich the embeddings with critical scheduling insights. The decoder comprises two components: a memory network that generates job-specific states using handcrafted contextual features and multi-head attention to model inter-job dependencies, and a classifier network that combines the embeddings and job states to compute selection probabilities through a feedforward neural network and softmax activation. The model uses a self-supervised training strategy called the Self-Labeling Improvement Method. This strategy generates multiple solutions for each instance, selects the one with the minimum makespan as a pseudo-label, and optimizes the model parameters using a cross-entropy loss. By iteratively learning from its own outputs, the model progressively refines its ability to produce high-quality solutions without relying on external optimality information.

\begin{table}[h]
\renewcommand{\thetable}{\arabic{table}}
 \centering

\caption{Deep Neural Networks models for JSSPs.}
\resizebox{13.5cm}{!}{
\begin{tabular}{l|p{5cm}|l}

\hline\noalign{\smallskip}
Study & NN Model & JSSP representation  \\
\noalign{\smallskip}\hline\noalign{\smallskip}
Zang et al \cite{95} &Hybrid deep convolution and back propagation network & GA-generated schedules\\
Lara-Cárdenas et al \cite{18} & Deep multilayer perceptron network& Constructive hyper-heuristic-generated schedules\\
Shao and Kim  \cite{88} &Long Short-Term Memory network& GA-generated schedules\\
Wang et al \cite{81} &Deep convolutional neural network& Constraints satisfaction problem\\
Juros et al  \cite{33} & Convolutional graph neural network& Bipartite Graph \\
Wang et al \cite{GNN1} &Graph neural network& Disjunctive Graph \\
Corsini et al \cite{GNN2} & Graph attention network& Disjunctive Graph \\
\noalign{\smallskip}\hline
%%\label{tab:01}

\end{tabular}
}
\label{tab dnn}

\end{table}

Deep neural networks have been applied both as standalone solvers and in combination with traditional optimization techniques to tackle the JSSP. When used independently, DNNs are capable of learning intricate patterns and extracting features from problem instances to provide solutions directly, such as predicting makespan or generating scheduling priorities. They analyze the detailed features of scheduling data to predict optimal solutions or classify priorities, often achieving significant improvements in computational efficiency. In hybrid approaches, DNNs complement traditional optimization techniques by enhancing the efficiency and decision-making process. These models improve classical methods by learning heuristic rules or approximating computationally expensive processes (e.g., variable selection in branch-and-bound). A particularly innovative approach involves representing JSSP instances as graphs, where nodes represent jobs, operations, or machines, and edges capture dependencies or precedence constraints. Graph neural networks are well-suited for processing such representations, as they leverage the graph structure to encode relational information and dependencies between elements. Table \ref{tab dnn} provides a consolidated overview of the deep learning networks discussed for the JSSP.

\subsection{Reinforcement Learning }
\begin{figure}[h]
  \centering
  \resizebox{8cm}{3cm}{
\begin{tikzpicture}[->,node distance=1.3cm,>=stealth',bend angle=40,auto,
  place/.style={rectangle,thick,draw=red!75,fill=gray!5,minimum size=10mm},
  red place/.style={place,draw=red!75,fill=red!20}
  every label/.style={red},
  every node/.style={scale=.6},
  dots/.style={fill=black,circle,inner sep=2pt},
  initial text={}]
 \node [ place,label={[shift={(0.4,-1.9)}]}] (l1-1)  {$Agent$};
  \node [place,right=2cm of l1-1] (l1-2) {$Environment$};

  \path (l1-1) %edge [in=160,out=190,loop,align=center] node[left]{vp\\$x:=0$} (l1-1)
           %edge [in=240,out=210,loop,align=center] node[below,xshift=-7mm]{vs\\
          % $x < 850$\\$x:=0$} (l1-1)
           %edge [in=140, out=110,loop,align=center] node[above]{ap\\$x \ge 850$
           %\\$x:=0$} (l1-1)
    (l1-1) edge node[above]{ $(s_t, a_t$)} (l1-2)
    (l1-2) edge [bend right] node[above,align=center]{ $r_t$} (l1-1)
    edge [bend left] node[below,align=center]{$ s_{t+1}$} (l1-1)
    ;
\end{tikzpicture}
}
\caption{Reinforcement learning one time step.}\label{fig3}
\end{figure}
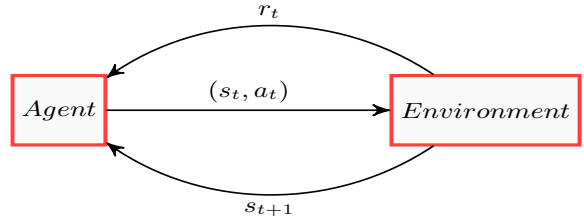

As noted by Kaelbling et al. \cite{44}, reinforcement learning (RL) is "a way of programming agents by reward and punishment without needing to specify how the task is to be achieved." RL involves a conversation between an agent and its environment. At time \textit{t}, the environment reveals itself to the agent in the form of a state \textit{s}. To change this state, the agent takes an action \textit{a}. This action generates a reinforcement signal, called \textit{r}. The agent receives this signal and transitions to the next state \textit{$s^{'}$}. The agent's task is to learn a policy $\pi$ that maps states to actions, $\pi(s) = a$, in order to maximize the expected reward. In Figure \ref{fig3}, we describe the RL conversation between the agent and its environment.

Formally, basic RL is modeled as a Markov Decision Process (MDP), which consists of: \begin{itemize} \item $S$: the set of environment states; \item $A$: the set of actions; \item $R$: the reward function, $R: S \times A \rightarrow \mathbb{R}$; \item $T$: the transition function, defined as: \begin{equation} T(s, a, s^{'}) = P_r(s_{t+1} = s^{'} | s_t = s, a_t = a) \end{equation} where $P_r$ is the probability of transitioning to $s_{t+1}$ after executing action $a_t$ in state $s_t$. \end{itemize}

As in equation (\ref{eq:01}), the value function $V^{\pi}(s)$ of a state \textit{s} is the expected return for the agent when starting in state \textit{s} and following the policy $\pi$. $R(s,\pi(s))$ is the reward received for taking the action \textit{a} while in state \textit{s} and transitioning to the state \textit{$s^{'}$}. The discount factor $\gamma \in [0, 1]$ balances the importance between immediate rewards ($\gamma = 0$) and long-term rewards ($\gamma = 1$).

\begin{equation} V^{\pi}(s) = R(s,\pi(s)) + \gamma \left( \sum_{s^{'} \in S} T(s, \pi(s), s^{'}) V^{\pi}(s^{'}) \right). \label{eq:01} \end{equation}

To solve the MDP, the goal is to find an optimal policy. As shown in equation (\ref{eq:02}), $\pi^{*}(s)$ depends on the transition and reward functions.

\begin{equation} \pi^{*}(s) = \underset{a \in A}{\text{argmax}} \left[ R(s,a) + \gamma \sum_{s^{'} \in S} T(s, a, s^{'}) V(s^{'}) \right] \label{eq:02} \end{equation}

If the transition function $T$ and the reward function $R$ are known, $\pi^{*}$ can be found through iterative methods such as policy iteration or value iteration.

\begin{itemize} \item {\textbf{Policy iteration:} As described in Algorithm \ref{Algo:01}, policy iteration performs two operations: policy evaluation and policy improvement. It starts with a random policy $\pi$ at time $t = 0$ and performs the policy evaluation. For each state $s \in S$, it calculates the value function under the initialized policy. Then, it updates the value function until the change "$\Delta \leftarrow \max(\Delta, |v - V(s)|)$" between the current and old value is sufficiently small. Here, $\Delta$ is initialized to 0, and $\epsilon$ is a predefined error threshold. Based on the value function computed during the policy evaluation, the policy improvement updates the policy itself for every state by calculating the reward for each action until the optimal policy is found.} \end{itemize}

\begin{algorithm}[H]
\SetAlgoLined
\begin{flushleft}\textbf{Initialization}\\
 $v(s) \in \mathbf{R}$ and $\pi(s) \in A(s)$ arbitrarily for all $s \in S.$
\end{flushleft}
\begin{flushleft} 

  \textbf{Policy evaluation}\\
  \begin{flushleft}
\Repeat{$\Delta < \epsilon$( a small positive number)}
{$\Delta \leftarrow 0$\\
for each $s \in S$\\
$v \in V(s)$\\
$V(s) \leftarrow$ $ \underset{a \in A}{Max} [R(s,a)+\gamma\sum_{s^{'}\in S}T(s,a,s^{'})V(s^{'})] $\\
$\Delta \leftarrow Max(\Delta, |v-V(s)|)$
 }\end{flushleft}
 
\textbf{Policy improvement}\\
 \begin{flushleft}
 Policy-stable $\leftarrow $true\\
 for each $ s\in S$\\
    	\quad  $a\leftarrow \pi(S)$\\
    	\quad   $\pi(s)\leftarrow  \underset{a \in A}{argmax}[R(s,a)+\gamma\sum_{s^{'}\in S}T(s,a,s^{'})V(s^{'})]$\\
    	\quad  If $a \neq \pi(s)$, Then  Policy-stable $\leftarrow $ flase\\
If  Policy-stable, Then Stop and return $V$ and $\pi$; else go to \textbf{Policy evaluation}  
 \end{flushleft}
\end{flushleft}
 \caption{Policy iteration Algorithm}
 \label{Algo:01}
\end{algorithm}

\begin{itemize} \item {\textbf{Value iteration:} Value iteration (Algorithm \ref{Algo:02}) combines policy evaluation and policy improvement, using the Bellman equation \cite{58} as an update rule. Similar to policy iteration, it starts with an arbitrary initialization. For each $s \in S$, it computes the value function and continues updating until the change between the current and the old value function, denoted as "$\Delta$", becomes smaller than the predefined error threshold $\epsilon$. It then returns the optimal policy $\pi^{*}$.} \end{itemize}

\begin{algorithm}[H]
\SetAlgoLined
\begin{flushleft}
\item
\KwIn{An MDP model}
\end{flushleft}
\begin{flushleft}
 \textbf{Initialization} \\
  $ \forall s \in S: V(s)$ arbitrarily \\
\end{flushleft}
\begin{flushleft}
\Repeat{$\Delta < \epsilon$ (a small positive number)}{
    $\Delta \leftarrow 0$ \\
    for each $s \in S$ \\
    $v \leftarrow V(s)$ \\
    $V(s) \leftarrow \underset{a \in A}{\text{Max}} \left[ R(s,a) + \gamma \sum_{s' \in S} T(s,a,s') V(s') \right] $ \\
    $\Delta \leftarrow \text{Max} (\Delta, |v - V(s)|)$
}
\end{flushleft}
\textbf{Output} $\pi(s) = \underset{a \in A}{\text{argmax}} \left[ R(s,a) + \gamma \sum_{s' \in S} T(s,a,s') V(s') \right]$
\caption{Value Iteration Algorithm}
\label{Algo:02}
\end{algorithm}

Model-free and Model-based are foundational RL approaches. Model-free methods learn the policy directly from the rewards without explicitly learning the transition function. In contrast, Model-based approaches involve estimating the transition function for all states, allowing the agent to plan its actions using the learned model of the environment. Several RL approaches exist in the literature. Among them is Q-learning, a well-known model-free RL approach \cite{Watkins1992}. Q-learning associates for each state-action pair $(s, a)$ a Q-value (eq. (\ref{eq13})) and stores it in a table. The agent gains experience by trial and error through the execution of actions and updates all its Q-values using the following equation:  \begin{equation}
         Q(s,a) \leftarrow Q(s,a)+\alpha [r+ \gamma \underset{a'}{max} Q(s^{'},a^{'})-Q(s,a)]
         \label{eq13}
 \end{equation}
 $\alpha \in [0,1]$, is learning rate and \textit{r} is the reward or penalty resulting from taking action \textit{a} in state \textit{s}.

For the JSSP, various representations of Q-learning are possible, and it can be utilized as a method for selecting dispatching rules. Adin and Oztemel \cite{45} proposed a Q-learning algorithm to train a reinforcement learning agent that interacts with a JSSP environment. The agent selects appropriate dispatching rules from (Shortest processing time, cost over time and critical ratio). Similarly, Wang and Usher \cite{91} design a Q-learning agent to determine the most suitable operation at each iteration. The experimental results demonstrated that Q-learning provides better solution quality compared to standalone dispatching rules. Fonseca et al. \cite{92} propose a basic Q-learning algorithm to find a sequence of operations that minimizes the makespan of the job-shop.

As the number of state-action pairs increases, Q-learning demands greater computational effort to calculate the value function. Gabel and Riedmiller, \cite{78} propose approximating the value function using a multilayer perceptron neural network. Moreover, the JSSP is represented as a Multi-Agent Markov Decision Process (MMDP), where each machine in the problem is associated with a Q-learning agent. Each agent independently learns its dispatching rules and refines its behavioral policy within a multi-agent learning algorithm designed to ensure the independent learning of all agents. Additionally, Yailen and Nowe \cite{89} introduce a decentralized MMDP model with changing action sets to represent the JSSP. In this model, each machine is treated as a Q-learning agent that learns the processing order of operations.

The above reinforcement learning approaches have shown the great success of RL methods for solving different variants of job-shop scheduling problems, summarized in Table \ref{tab RL}.  Despite this success, the learning algorithms are suffering from a lack of generalization, especially the Q-learning approach. This is mainly due to the huge number of states to visit in training and the large storage space needed in the q-table.  The ideal way is to perform the learning for a small number of training states and generalize this experience to new similar states. This leads to Deep Reinforcement Learning (DRL).   Although DRL method is effective for solving JSSP, there are still deficiencies in state representation, action space definition, and reward function design, which make it difficult for the agent to learn effective policy

\begin{table}[h]
\renewcommand{\thetable}{\arabic{table}}
 \centering

\caption{Reinforcement learning approaches for JSSPs.}
%\resizebox{13.5cm}{!}{
\begin{tabular}{l|l|p{4cm}}

\hline\noalign{\smallskip}
Study & RL environnement & RL approach \\
\noalign{\smallskip}\hline\noalign{\smallskip}
Adin and Oztemel \cite{45} & Single MDP agent MDP & Q-learning \\
Wang and Usherl \cite{91} &   Single MDP agent  & Q-learning  \\
Fonseca et al.  \cite{92} &   Single MDP agent  & Q-learning  \\
Gabel and Riedmiller \cite{78} &Multi-Agent MDP & Q-learning based multilayer perceptron network\\
Yailen and Nowe  \cite{89} & Multi-Agent decentralized MDP & Q-learning \\

\noalign{\smallskip}
\hline
%%\label{tab:01}

\end{tabular}
%}
\label{tab RL}

\end{table}

\subsubsection{Deep Reinforcement Learning}

The Deep Reinforcement Learning (DRL) methods \cite{83} represent reinforcement learning techniques that integrate deep learning technologies. With the widespread application of deep learning, most RL algorithms now incorporate these methods. Specifically, DRL primarily uses deep neural networks to approximate two variables: value functions and policies. 

Bruno et al. \cite{6} highlight the potential of DRL as a modern method for solving scheduling problems. They proposed a DRL architecture to solve the JSSP using the Deep Q-Network algorithm (DQN).  Lin et al. \cite{11} employed a DQN to solve the JSPP within a smart factory framework called Multi-class Deep Q-Learning (MDQL). The idea was to train a multi-layered network to decide a dispatching rule for each machine from the following set: FIFO, SPT, LPT, MOPNR, LOPT, SQN, and LQN. This work was the first to use multiple dispatching rules, rather than a single dispatching rule, for solving the JSSP. The approach was compared with random job dispatching. Various parameter settings (such as the number of neurons in the hidden layer, the number of epochs, the e-greedy policy, the learning rate for the update function, and the number of jobs and machines) were reported to analyze their effect on the convergence of the model. The results demonstrated the strong performance of the proposed model.

Liu et al. \cite{10} introduce a pioneering DRL approach to solve the JSSP in both dynamic and static environments. The JSSP is modeled as a multi-agent Markov Decision Process. The DRL agent interacts with its environment using two neural networks. Each network includes a fully connected convolutional layer with a nonlinear activation function. The first network, referred to as the critic network, evaluates the state and approximates the Q-values for state-action pairs. The second network, known as the actor network, determines the agent’s behavior by leveraging the critic network's approximations. To train the model in a dynamic environment, they employed a parallel training method that combines asynchronous updates with an adapted DQN algorithm called Deep Deterministic Policy Gradient (DDPG). The proposed approach was evaluated using OR-library instances and compared against existing dispatching rules, meta-heuristics, generic multi-agent Q-learning \cite{89}, and optimal solutions. The results demonstrated the efficiency and effectiveness of the DRL agent in handling the JSSP.

 The strength of DRL is closely linked to the neural network used, making the choice of network critical. Since our problem is consistently represented as a graph, it becomes even more interesting for the neural network to learn directly from this representation. Zhang et al. \cite{12} developed a DRL approach for classical JSSPs based on the Graph Isomorphism Network (GIN). GIN is a new variant of GNNs with enhanced discrimination power, designed to operate on undirected graphs. As such, GIN supports the structure of the JSSP disjunctive graph. The authors modify the disjunctive graph to create a new scheduling graph, which is then used by GIN to learn a parameterized stochastic policy. This policy is trained using the Proximal Policy Optimization (PPO) algorithm \cite{36}, a policy gradient-based reinforcement learning algorithm. The model was tested on small, medium, and large-scale JSSP instances. The learned dispatching rule outperformed traditional methods. Similarly, Hameed and Schwung \cite{49} proposed a DRL approach based on GNN for the JSSP. In contrast to \cite{12}, this study uses a multi-layer perceptron  as a GNN, which provides a suitable representation of the problem without altering the scheduling graph. In this model, the neurons of the neural network represent jobs and machines, while the edges define the connections between them. Instead of using a single agent, this approach integrates the model into a multi-agent system. Each agent observes its state from a subset of the entire state space, selects an action from its available set, and receives an individual reward. Each agent optimizes its policy using the same learning algorithm (PPO) as in \cite{12}. The model is applied to solve real-world JSSPs, such as robot manufacturing cells, and several benchmark instances of the problem. A limitation of this approach is the lack of communication between agents. Each agent reacts individually based on its subset of states, without observing the entire state space or collaborating with other agents.

Park et al. \cite{Park} propose a DRL approach using a GNN model to capture features from the disjunctive graph representation of the JSSP. These features are processed through the GNN layers, where message-passing mechanisms aggregate information from neighboring nodes and edges. The learned graph embeddings are then integrated into a reinforcement learning framework based on the actor-critic architecture. The actor network leverages these embeddings to output a stochastic scheduling policy, determining the next operation for each machine. Simultaneously, the critic network assesses the quality of the current state-action pair by estimating its value function. To train the scheduling policy, they employed the PPO method, which optimizes the parameters of both networks by iteratively refining the policy to minimize the makespan. Their approach outperformed outperforming traditional rule-based and heuristic methods and demonstrated strong generalization across benchmarks of varying sizes.

Park et al. \cite{Park1} present the JSSP as a multi-agent problem, where machines are represented as agents. Each agent is associated with a Type-Aware Graph Attention (TGA) component, which belongs to the family of GNNs. The model is similar to the one proposed in \cite{Park}, with the key distinction being the use of the TGA network to capture features from the disjunctive graph. The TGA network combines the strengths of GNNs and attention mechanisms with type-specific awareness, allowing it to efficiently manage the complexity and scale of large JSSP instances. Its modular and scalable design, coupled with integration into RL frameworks, ensures robust performance across a range of problem sizes and constraints.

Liu et al. \cite{Liu} formulate JSSP instances as a machine-job graph, represented by three matrices: disjunctive, weight, and state matrices and graph embedding layer based on a Dueling Double Deep Q-network. Double Deep Q-Network (DDQN) \cite{DDQN} is an improved version of the DQN method, designed to overcome the overestimation of Q-values that occurs when the algorithm uses the same Q-values to both choose and evaluate actions. This is achieved by introducing two separate neural networks: the Q-network and the target Q-network. Instead of using a single network for both action selection and evaluation, DDQN uses two distinct networks: one for selecting actions and another for evaluating them. This reduces Q-value overestimation and improves the stability of the learning process. On the other hand, Dueling DQN \cite{DueDQN} addresses the issue of distinguishing between the value of a state and the advantage of taking an action, which makes learning more challenging. Dueling DQN modifies the network architecture by separating the estimation of the state value and the advantage of each action. This allows for better separation of information related to the state and the actions, thereby speeding up and improving the learning process. \cite{Liu} combines Double and Dueling DQN to tackle the JSSP. The model demonstrates its strength against standard dispatching rules. 

Liao et al. \cite{Liao} focus on the JSSP disjunctive graph and explore a hierarchical reinforcement learning method based on a DQN network with a two-level hierarchical structure. The High-Level Controller receives the current global state and outputs a sequence of operations to be scheduled. The Low-Level Controller receives the decision from the High-Level Controller and uses a Q-network to learn the optimal scheduling of operations on machines. 

Hameed and Schwung \cite{Ham} present the JSSP as a bipartite graph with two sets of vertices: the set of machine buffers and their corresponding set of machines. The edges are calculated dynamically at each time step based on the jobs available in the environment and their remaining operation sequences. They propose different GNN architectures to capture the features of the bipartite graph, along with a message-passing mechanism between a machine and its connected machine buffers, or vice versa. Furthermore, they adopt PPO agents to assign operations to each machine. While, Chen et al. \cite{chen} propose a DRL approach based a GNN network that integrates attention mechanism and disjunctive graph embedding to solve the JSSP. The Network relies on REINFORCE method \cite{Williams1992} to train the model. 

Lee et al. \cite{Lee} propose a DRL approach to tackle combinatorial optimization problems, with a focus on the JSSP. The approach combines a Transformer-based neural network with an attention mechanism. By focusing on the most relevant parts of the scheduling problem at each step, the model is able to learn a more effective and optimal scheduling policy. The training involves standard RL techniques with a policy gradient method to optimize the scheduling policy and minimize the makespan. The experiments demonstrate that the model outperforms standard dispatching rules.

We conclude our study with the original work of Zhang et al. \cite{Zhang}. They highlighted the potential of using DRL and graph-based techniques to guide neighborhood-based improvement heuristics, marking a significant advancement in scheduling optimization. They introduced a novel graph embedding scheme that combines information from the disjunctive graph topology with the heterogeneity of neighboring nodes. This embedding scheme enhances the model's ability to comprehend the structure of the JSSP. Additionally, they designed a message-passing mechanism to evaluate batches of solutions more efficiently, thereby accelerating the heuristic improvement process. This approach represents a promising new direction for applying DRL to improvement heuristics in scheduling. Table \ref{tab DRL} summarizes the DRLs approaches used to solve the JSSP. 
\begin{table}[h!]
\renewcommand{\thetable}{\arabic{table}}
 \centering

\caption{Deep Reinforcement learning approaches for JSSPs.}
\resizebox{13.5cm}{!}{
\begin{tabular}{l|l|p{4cm}|l}

Study & RL environnement & DRL Nework &Learning method \\ 
\noalign{\smallskip}\hline\noalign{\smallskip}
Bruno et al. \cite{6}& Single agent& Multi-layered network & Deep Q-Network \\
Lin et al. \cite{11}& Single agent & Multi-layered network & Multi-class Deep Q-Learning\\
Liu et al. \cite{10}& Multi-agent & Fully connected convolutional netwok  & Deep Deterministic Policy Gradient\\

Zhang et al. \cite{12}& Single agent & Graph Isomorphism Network &Proximal Policy Optimization \\

Hameed and Schwung \cite{49}& Multi-agent & Graph neural network &Proximal Policy Optimization \\
Park et al. \cite{Park}& Single agent& Graph network based message-passing mechanism &Proximal Policy Optimization \\
Park et al. \cite{Park1}& Multi-agent & Type-Aware Graph Attention& Proximal Policy Optimization \\
Liu et al. \cite{Liu}& Single agent &Graph neural network & Double and Dueling DQN\\
Liao et al. \cite{Liao}& Single agent& Graph neural network & Deep Q-Network\\
Hameed and Schwung \cite{Ham}& Single agent & Graph network based message-passing mechanism&Proximal Policy Optimization  \\
Chen et al. \cite{chen}& Single agent & Graph neural network & REINFORCE method\\

Lee et al. \cite{Lee}& Single agent&Transformer-based neural network with an attention mechanism&Policy gradient method \\
Zhang et al. \cite{Zhang}& Single agent & Graph network based message-passing mechanism& Policy gradient method \\

\hline
%%\label{tab:01}

\end{tabular}
}
\label{tab DRL}

\end{table}
\section{Discussion and Future Work}\label{sec:04}
According to the papers referenced above, the application of neural networks and reinforcement learning to JSSPs dates back over twenty years. In this study, we have traced the chronological evolution of these machine learning-based approaches.

Since the inception of neural network-based approaches for solving job shop scheduling problems, research has progressed through three distinct stages. The first stage marked the earliest attempts to address JSSPs using Hopfield NNs and enhanced constraint satisfaction NNs. These methods were limited by issues such as constraint violations and an inability to handle large-scale problem instances. Consequently, their performance was modest compared to other contemporary approaches. The second stage introduced methods that can be categorized into two main types:Integration with classical optimization methods: NNs were employed to enhance traditional optimization techniques, either by providing high-quality initial solutions for population-based methods or by dynamically adjusting heuristic parameters. Supervised learning from optimization algorithms-generated solutions:
In this category, NNs were trained using solutions derived from the optmization algorithms, incorporating a variety of dispatching rules. These NNs captured essential scheduling features and predicted priorities for activities or the appropriate dispatching rules for machines. The trained models were then applied to solve new JSSP instances not encountered during training. While these models demonstrated the ability to generalize their predictions to unseen instances, they faced notable limitations. The results were superior to the traditional training methods used to create them; however, their generalization capacity diminished significantly when the sizes of the training and testing instances differed. Furthermore, training times were lengthy, and the accuracy was often insufficient. Both the first and second-stage approaches relied on shallow learning techniques. In contrast, the third stage is characterized by the adoption of deep learning such graph-based NNs. Deep NNs with additional hidden layers reduce errors and improve accuracy. Moreover, their architecture enables the integration of diverse types of layers within a single model, enhancing both precision and generalization across various JSSP variants. Despite these advancements, deep learning methods come with challenges. Training times increase significantly as the number of hidden layers grows, due to the complexity of the network architecture. Additionally, there is currently no theoretical framework for determining the optimal number of hidden layers or neurons specifically for JSSPs, which complicates the model design process.

Concerning RL-based methods, they have evolved through two main stages. The first stage focuses on the initial attempts to adapt reinforcement learning to solve JSSPs using the MDP model and the Q-learning algorithm. The goal is to enable the agent to learn its own scheduling decisions. As anticipated, the results showed that the RL agent consistently learns the appropriate dispatching rule. However, these techniques still face the same limitations as traditional AI methods previously used to solve JSSPs. As the complexity of the problem increases, so does the state space of the model. Consequently, Q-learning becomes increasingly time-consuming and demands large amounts of memory. To address these issues, multi-agent RL approaches have been proposed. These systems can be adapted in two ways, depending on the role of the agent. In the first approach, the agent represents the job, while in the second, the agent represents the machine. In both cases, the learning time is reduced compared to the first-stage method. This reduction in time is a result of decomposing the search space across multiple agents. However, when agents work independently without communication, the lack of shared information about the global problem state negatively impacts the quality of the learned policies. Furthermore, the choice of each agent’s local reward, along with the system’s global reward, is a critical factor in the learning process. Introducing communication between agents, although beneficial, requires significant computational resources. As a result, these multi-agent RL techniques are less suited for JSSPs, particularly in larger and more complex environments. The second-stage approaches are presented by deep reinforcement learning models, which have proven to be a valuable optimization tool. As explored in various studies, DRL models can handle the complexity and dynamic nature of JSSPs by using deep neural networks to approximate value functions and policies. A notable advantage of DRL is its ability to learn directly from the environment, adjusting its strategies based on the feedback it receives, which can significantly improve the scheduling process. One of the most compelling aspects of DRL in the context of JSSP is its ability to leverage graph-based representations, particularly with the integration of Graph Networks. By framing the JSSP as a graph, the model can exploit the rich structural information inherent in the problem, leading to more efficient learning and better solutions. Graph representations in DRL models improve learning efficiency by capturing dependencies and relationships between components, allowing for faster convergence and better performance, particularly in large-scale problems. Graph-based models also scale well, handling the growing complexity of larger JSSPs. Additionally, they enhance generalization, enabling DRL models to apply learned knowledge across various problem instances. Hierarchical and multi-agent approaches leverage graph representations to improve scalability and coordination among agents. Furthermore, graph embeddings optimize action selection and policy refinement, aiding in better scheduling decisions. Finally, these models offer customization and flexibility, adapting to specific constraints or objectives in real-world scheduling scenarios.

Recently, there has been a focus on integrating deep learning methods with traditional approaches to address the complexities of the Job Shop Scheduling Problem. By leveraging Deep Reinforcement Learning, we aim to enhance the performance of existing heuristics and improve scheduling decisions. Additionally, we will explore the challenging variant of the JSSP with Blocking Constraints (BJSSP), a problem that has yet to be fully addressed using learning-based methods. Our goal is to develop novel DRL-based techniques that can efficiently handle the increased complexity introduced by blocking constraints, providing new solutions for more realistic and large-scale scheduling problems.

\section{Conclusion}
\label{sec:05}

Machine learning techniques have become an important focus for researchers in the field of combinatorial optimisation, particularly for Job Shop Scheduling Problems. Among the various methods within machine learning, neural networks and reinforcement learning have been the most prominent in solving JSSPs. The first application of neural networks to JSSPs dates back to 1988, and since then there have been numerous improvements and innovations, as highlighted in this survey. While these approaches have shown some success, they have also faced limitations, particularly in terms of storage capacity and computational time. Newer approaches, such as deep neural networks, have emerged to address these challenges. Reinforcement learning, on the other hand, is relatively new compared to traditional methods. Its application to JSSPs has primarily involved various forms of Markov decision processes and multi-agent systems. Reinforcement learning algorithms have been widely used in this area, having been adapted and tailored to the constraints of JSSPs. While the results demonstrate the strength and effectiveness of these methods, they often suffer from a lack of generalisation and convergence for large problems. This is due to the fact that the state-action space in reinforcement learning grows exponentially, making naive exploration infeasible and requiring specialised techniques such as: incorporating neural networks, pruning/compressing the state-action space, and using hierarchical RL representation, which remains an active area of research.

\end{document}